\documentclass[showpacs,eqsecnum,aps,superscriptaddress]{revtex4}

\usepackage{graphicx}%
\usepackage{dcolumn}
\usepackage{amsmath}

\newcommand{\rst}{\rho^*}

\newcommand{\pa}[1]{\partial_#1}

\begin{document}

\preprint{??}

\title[Short Title]{Conformally flat Smoothed Particle Hydrodynamics\\ Application to Neutron Star Mergers}

\author{Roland Oechslin}
\email{Roland.Oechslin@unibas.ch}
\affiliation{Department of Physics and Astronomy, University of Basel, Klingelbergstr. 82, 4056 Basel, Switzerland}
\author{Stephan Rosswog}
\affiliation{Department of Physics and Astronomy, University of Leicester, LE1 7RH, Leicester, UK}
\author{Friederich-Karl Thielemann}
\affiliation{Department of Physics and Astronomy, University of Basel, Klingelbergstr. 82, 4056 Basel, Switzerland}

\date{\today}
\begin{abstract}
We present a new 3D SPH code which solves the general relativistic field+hydrodynamics equations in the conformally flat approximation. Several test cases are considered to test different aspects of the code. We finally apply then the code to the coalescence of a neutron star binary system. The neutron stars are modeled by a polytropic equation of state (EoS) with adiabatic indices $\Gamma=2.0$, $\Gamma=2.6$ and $\Gamma=3.0$. We calculate the gravitational wave signals, luminosities and frequency spectra by employing the quadrupole approximation for emission and back reaction in the slow motion limit. In addition, we consider the amount of ejected mass.

\end{abstract}


\maketitle

\section{Introduction}
Coalescing binary neutron stars (NS) are expected to be one of the most powerful sources of gravitational radiation for detectors like LIGO \cite{ligo}, VIRGO \cite{virgo}, GEO600 \cite{geo} and TAMA \cite{tama}. In contrast to electromagnetic radiation, gravitational waves (GW) are hardly perturbed during their journey from their emission site to their detector. Since the GW emission is sensitive to mass distributions it provides additional information on the astrophysical object that is not available from electromagnetic radiation. To extract a signal from the noisy output of the interferometers, precise theoretical templates of the waveforms are needed.\\
Merging NS are also of interest to the nuclear astrophysics community since they are a promising site for the formation of the heaviest elements via the rapid neutron capture process (r-process). The decompression of neutron-rich material through tidal disruption has first been proposed as r-process-scenario by Lattimer \cite{lattimer74,lattimer76} and has been later investigated by several groups \cite{Meyer,symbalisty,eichler,rosswog99,rosswog00,freiburghaus}. Apart from that, the coalescence of two NS is at the heart of several models for the still poorly understood Gamma-Ray Bursts (e.g. \cite{Duncan, kluzniak}).\\
Since 1990, various groups have performed hydrodynamical calculations of merging NS. The first work, using Newtonian gravity and a polytropic EoS was done by Oohara \& Nakamura (\cite{oohara97} and references therein) using an Eulerian finite difference scheme. The problem has been further explored by Davies et al. \cite{davies}, Zhuge \cite{zhuge,zhuge2} and Rasio \& Shapiro (\cite{Rasio} and references therein) using smoothed particle hydrodynamics. Improvements towards a full general relativistic implementation were achieved by Ayal et al. \cite{shai} and Faber et al. \cite{faber,faber2} using the Post-Newtonian approximation and Wilson et al. \cite{wilson} using the conformally flat approximation. The first fully relativistic calculations have been carried out by Shibata \& Ury$\bar{\text{u}}$ \cite{shibataGR, shibatauryu, shibataGRhires}. The microphysical aspects have been investigated by Ruffert et al. (\cite{ruffert96,max} and references therein) and Rosswog et al. (\cite{rossi2001} and references therein), both using a realistic Nuclear EoS. For a review on the topic, see \cite{rasiorev}.\\
The merger phase of a Neutron star binary represents only the last phase of a evolution sequence beginning with a long inspiral phase during which the binary evolves on nearly stationary orbits. As energy is radiated away due to GW, the binary separation slowly decreases down to the innermost stable circular orbit (ISCO) where hydrodynamical instabilities due to tidal forces set in. During the following merger phase, the two stars are disrupted and a new central object, either a NS or a black hole (BH) is formed in the center. This object finally settles down to equilibrium in a ring-down phase. Since the time scales for each phase are drastically different from each other, different computational approaches have to be chosen for each phase.\\
The inspiral phase can be approximated to high precision by a pointmass-Post-Newtonian (PN) approximation \cite{blanchetpointmass,blanchetpointmass2}. This approximation is valid as long as finite-size and strong-field effects are negligible. If this is not the case, i.e. for very close orbits, a more exact calculation is provided by the quasi-equilibrium approximation \cite{duezQE} which is based on quasi-equilibrium sequences in the conformally flat approximation. Very close to the ISCO this approximation breaks also down and dynamical calculations have to be carried out. The outcome of the merger does not only depend of the GR effects but also on intrinsic properties of the neutron stars (NS) like the equation of state (EoS). The appropriate tool to treat this problem is a fully relativistic magnetohydrodynamic simulation including all the microphysics involved. First successes have recently been reported in implementing the GR component of the problem \cite{shibataGR, shibatauryu, shibataGRhires}. However, these calculations still remain very complicated due to the nature of the non-linearity of the field equations and the enormous demand in computational resources.\\
Beside the PN approximation to GR there exists also the approximation using the conformally flat condition (CFC). Both are of about the same computational complexity. The PN approximation, which is a weak-field, slow-motion expansion of the field equations does not describe the problem adequately \cite{shai,faber} in first order (1PN) since the expansion parameters within the NS have values of $\sim 0.4$ and the convergence is generally rather slow. It should be mentioned, however, that there has been progress in the binary-pointmass case \cite{damour98} with Pad\'e-approximants. On the other side, the CFC approximation can also describe strong field scenarios in special cases. The static Oppenheimer-Volkoff (OV) profile is reproduced exactly, uniformly rotating polytropes are described up to a precision of $\simeq$5\% \cite{cook}. For a binary near the ISCO, \cite{shibatahij} estimate the magnitude of the non-diagonal spatial metric components (which are neglected by the CFC approximation) to $\simeq0.02$ for a compactness of $M/R=0.14$ and $\simeq0.05$ for $M/R=0.19$, their influence for the angular velocity $\omega$ and the angular momentum $J$ to $\mathcal{O}(10^{-2})$  for the density $\rho$, the conformal factor $\psi$ and the mass $M$ to $\mathcal{O}(10^{-3})$. This systematical error will certainly increase in the merger phase, but the quality of the CFC-approximation is still not fully explored yet.\\
The first hydrodynamical calculations using the CFC approximation have been carried out by Wilson \cite{wilson} with the irritating result of a possible collapse of both NS prior to merging. After correcting an error in the original formulation \cite{flanagan} the method is well established today to investigate quasi-equilibrium configurations. New, independent hydrodynamical calculations have never been carried out since then. Doing so, such calculations could explore the applicability of the CFC approximation during the merger phase and eventually establish it as an economical approximation method to GR. As there exist other, much larger uncertainties on the numerical and microphysical side of the problem, a robust and economical simulation code based on a well-explored approximative method could be a step ahead.\\
With this new code, using the CFC approximation, we therefore follow two goals: On one side, a fast and stable code to simulate compact objects is made available, on the other side, we want to explore the applicability of the CFC approximation during the merger phase by comparing our simulations to full GR results.\\
The paper is organized as follows. We present in section \ref{sec:numerics} the numerical method with the adaption of the smoothed particle hydrodynamics (SPH) method to the CFC approximation. In section \ref{sec:tests} we test and calibrate our code by applying it to two test scenarios that have previously been investigated with different methods. The central part of the paper, the simulation of the dynamical coalescence of two neutron stars is presented in section \ref{sec:merger}. Finally, we conclude the paper in section \ref{sec:conclusion}.

\section{Numerical Method}
\label{sec:numerics}

We use a Lagrangian particle method, the smoothed particle hydrodynamics (SPH), to implement the hydrodynamic part of the system of equations (\ref{conteqn}-\ref{energyeqn}) while the field equations (\ref{alphapsieqn}-\ref{chieqn}) are evaluated with a multigrid algorithm. The system of equations is summarized in appendix \ref{app:formalism}. For further references to SPH, see \cite{Monaghan,Benz}

\subsection{Basic SPH equations}
Since the coordinate conserved density $\rst$ in (\ref{stardens}) obeys a Newtonian-like continuity equation (\ref{conteqn}) we define $\rst$ in SPH by
\begin{equation}
\rst_a=\sum_b m_b W_{ab}
\end{equation}
where $m_b$ is the rest mass of particle $b$ and $W_{ab}=W(|\vec{r_a}-\vec{r_b}|,h)$ denotes the weight given by the standard spherical spline Kernel function $W(r,h)$.\\
The pressure gradient in the momentum equation (\ref{momentumeqn}) is calculated in a manner analogous to standard SPH, replacing $\rho$ by $\rst$,
\begin{equation}
\frac{1}{\rst_a}\vec{\nabla}p_a=-\sum_b m_b\left(\frac{p_b}{{\rst_b}^2}+\frac{p_a}{{\rst_a}^2}\right)\vec{\nabla}W_{ab}.
\end{equation}
The other terms in the momentum equation depend on values of the gravitational potentials and their derivatives. Their evaluation is described in the section (\ref{sec:fieldeval}).\\
The energy equation (\ref{energyeqn}) consists of Newtonian-like part which can be evaluated in the same manner as in standard SPH
\begin{eqnarray}
\frac{d}{dt}\epsilon_a&=&\frac{1}{2}\sum_b m_b\left(\frac{p_a}{\rho_a\rst_a}+\frac{p_b}{\rho_b\rst_b}\right)(\vec{v}_a-\vec{v}_b)\vec{\nabla}W_{ab}\\
&-&\frac{p_a}{\rho_a}\frac{d}{dt}\ln(\alpha u^0\psi^6)_a.
\end{eqnarray}
The total time derivative of $\ln(\alpha u^0\psi^6)_a$ in the second term, which arises due to the temporal change of the metric, is evaluated by second order finite differencing in time.
The hydrodynamic evolution is governed by the equation of state
\begin{equation}
p=(\Gamma-1)\rho\epsilon
\end{equation}
whereas initial data is generated by
\begin{equation}
\epsilon=\frac{\kappa}{\Gamma-1}\rho^{\Gamma-1}
\end{equation}
where $\kappa$ is the polytropic constant and $\Gamma$ the adiabatic index.\\
Since physical units enter only through the constant $\kappa$, the whole problem can be written in a non-dimensional way by the following transformation\cite{baumgarte}
\begin{eqnarray}
\bar{x}&:=&\kappa^{-n/2}x\nonumber\\
\bar{t}&:=&\kappa^{-n/2}t\nonumber\\
\bar{M}&:=&\kappa^{-n/2}M\label{kdimless}\\
\bar{\rho}&:=&\kappa^{n}\rho\nonumber\\
\bar{J}&:=&\kappa^{-n}J\nonumber\\
\bar{\Omega}&:=&\kappa^{n/2}\Omega\nonumber.
\end{eqnarray}
Therefore, $\kappa$ can be removed completely from the equations. Once results are obtained they can be rescaled to physical quantities via (\ref{kdimless}).{To integrate the system in time, a second order Runge-Kutta scheme with adaptive timestep is used. The timestep control mechanism is chosen as in \cite{lombardi} with $C_N=0.8$. Trial runs with smaller timesteps lead to similar results. The simulations are carried out on a standard 500 MHz Athlon workstation and last about three weeks, where the evaluation of the field equation in the nested grid setup takes most of the time.

\subsection{Artificial Viscosity}
We also include an artificial viscosity (AV) scheme to treat shocks.
The standard scheme proposed in \cite{MoGi} is known to give good results in shock situations but to introduce spurious viscosity forces in shear flows. We therefore choose a scheme \cite{MoMo} which assigns each particle $a$ its own time dependent viscosity coefficient $\alpha_a(t)$. The coefficient is determined by integrating the simple differential equation
\begin{equation}
\frac{d}{dt}\alpha_a=-\frac{\alpha_a-\alpha_{min}}{\tau_a}+S_a
\end{equation}
where $S_a=\max(0,-(\nabla v)_a)$.
The first term of the RHS causes $\alpha_a$ to decay on a timescale $\tau_a$ towards a minimum $\alpha_{min}$. The second term leads to a rapid increase of $\alpha_a$ only in case of a shock (i.e. $(\nabla v)_a$ largely negative).\\
To incorporate the AV scheme in a relativistic consistent manner we follow \cite{siegler} by adding the viscous pressure term $q_a$
\begin{equation}
q_a=
\begin{cases}
 \rho_a w_a(-\alpha_a c_a h_a(\nabla v)_a +\beta_a h_a^2(\nabla v)^2)_a & (\nabla v)_a>0\\
0 & \text{else}
\end{cases}
\end{equation}
to the physical fluid pressure,
where $\beta_a=2\alpha_a$, $w_a=1+p_a/\rho_a+\epsilon_a$ is the enthalpy, $h_a$ the smoothing length and $c_a$ is the speed of sound.\\
We directly use the SPH estimate for the velocity divergence
\begin{equation}
(\nabla v)_a\approx-\sum_b m_b(\vec{v}_a-\vec{v}_b)\vec{\nabla}_aW_ab
\end{equation}

\subsection{Evaluating the field equations}
\label{sec:fieldeval}

All six elliptic field equations for $\alpha\psi$, $\psi$, $B^i$ and $\chi$
(\ref{alphapsieqn},\ref{psieqn},\ref{beqn},\ref{chieqn}) are solved on an overlaid gravity grid via a multigrid solver (e.g. \cite{numrep}).\\
The transfer of all the quantities involved - hydrodynamical quantities as well as gravitational fields - between SPH particles and the grid points is done by assigning the SPH-interpolated values to the grid points and by interpolating the grid values to the SPH particles using a TSC scheme known from particle mesh codes \cite{hockney}.\\
To take into account the matter density distribution during the late stages of a neutron star merger (rapidly rotating central object plus low density material surrounding it), we additionally implement a nested grid algorithm which allows us to enlarge the computational domain while keeping the resolution in a selected subdomain. To extend the multigrid algorithm to the nested grid setup, we used the Multi Level Adaptive Technique (MLAT) (see e.g. \cite{Hess}).
Beyond these two grids, the potentials are extrapolated to the few particles which flow over the coarse grid. We use a quadrupole expansion around both individual stars for $\psi$
\begin{equation}
\psi\simeq1+\sum_{n=1}^2\left(\frac{M_\psi(n)}{r}+\frac{1}{2}Q_{\psi,ij}(n)\frac{x_i x_j}{r^5}\right)
\end{equation}
where 
\begin{eqnarray}
M_\psi(n)&=-&\int_{\text{star n}}S_\psi(x) d^3x\nonumber\\
Q_{\psi,ij}(n)&=-&\int_{\text{star n}}S_\psi(x) x_ix_jd^3x\nonumber
\end{eqnarray}
denote the monopole and the quadrupole of the potentials' source. The index $n$ runs over both stars since we expand around each star separately. A similar expression holds for $\alpha\psi$. These expansions follow from the field eqns. (\ref{psieqn}) and (\ref{alphapsieqn}).\\
For the shift-vector potentials, we make use of the imposed boundary conditions (\ref{approxbx}) - (\ref{approxchi})
\begin{eqnarray}
B^x(\mathbf{r}_a)&\simeq&\frac{b^x(\mathbf{r}_a)}{b^x(\mathbf{\bar{r}}_a)}B^x(\mathbf{\bar{r}}_a)\\
B^y(\mathbf{r}_a)&\simeq&\frac{b^y(\mathbf{r}_a)}{b^y(\mathbf{\bar{r}}_a)}B^y(\mathbf{\bar{r}}_a)\\
B^z(\mathbf{r}_a)&\simeq&\frac{b^z(\mathbf{r}_a)}{b^z(\mathbf{\bar{r}}_a)}B^z(\mathbf{\bar{r}}_a)\\
\chi(\mathbf{r}_a)&\simeq&\frac{c(\mathbf{r}_a)}{c(\mathbf{\bar{r}}_a)}\chi(\mathbf{\bar{r}}_a),
\end{eqnarray}
where $\mathbf{r}_a$ is the particle position of particle $a$ and $\bar{\mathbf{r}}_a$ its position projected onto the grid boundary.

\subsection{Backreaction Force and Gravitational Wave extraction}

To implement the radiation reaction and the GW extraction formulae (\ref{radreaction},\ref{GWextraction}) we consider the slow-motion quadrupole moment in SPH approximation 
\begin{eqnarray}
Q_{ij}&=&\text{STF}\left\{-2\sum_{ab}m_b W_{ab}\frac{S_{\psi b}}{\rst_b} x_b^i x_b^j\frac{m_a}{\rst_a}\right\}\nonumber\\
&=&\text{STF}\left\{-2\sum_b m_b \frac{S_{\psi b}}{\rst_b} x_b^i x_b^j\right\}
\end{eqnarray}
so that
\begin{equation}
\frac{d}{dt}{Q}_{ij}=\text{STF}\left\{-2\sum_a m_a \frac{S_{\psi a}}{\rst_a}x^i_a\left[v^j_a+\frac{d}{dt}\left(\ln\frac{S_{\psi a}}{\rst_a}\right)x^j_a\right]\right\}.
\end{equation}
To calculate gravitational wave signal and luminosity, we proceed as follows.
The second term in the brackets in much smaller than the first one \footnote{By approximating $x_j\approx r\sin(\omega t)$, $v_j\approx r\omega\cos(\omega t)$ and noticing $|\frac{d}{dt}\left(\ln\frac{S_{\psi b}}{\rst_b}\right)| \leq 5\times10^{-5}$ from trial runs, we get for the ratio of the amplitudes of the two contributions $|v^j_b|/|x^j_b|\simeq(r\omega)/(r|\frac{d}{dt}\left(\ln\frac{S_{\psi b}}{\rst_b}\right)|)\sim 10^{-3}$.} and therefore neglected. To obtain $\ddot Q_{ij}$ we take numerical derivatives of the above expression. This quantity is stored as a function of time and then smoothed and derivated numerically again in a postprocessing step to obtain the third time derivatives. With this data we are able to calculate the gravitational wave luminosity and the angular momentum loss without any further numerical approximation after the simulation. In addition, we determine the frequency spectrum using
\begin{equation}
\frac{dE}{df}(f)=\frac{\pi}{2}(4\pi r^2)f^2\langle|\tilde{h}_{+}(f)|^2+|\tilde{h}_{\times}(f)|^2\rangle
\end{equation}
where $\langle{h}_{+}\rangle$ and $\langle{h}_{\times}\rangle$ denote the angle-averaged waveforms \cite{zhuge}
\begin{eqnarray}
r^2\langle{h^2}_{+}\rangle&=&\frac{4}{15}(\ddot Q_{xx}-\ddot Q_{zz})^2+\frac{4}{15}(\ddot Q_{yy}-\ddot Q_{zz})^2\\
&+&\frac{1}{10}(\ddot Q_{xx}-\ddot Q_{yy})^2+\frac{14}{15}\ddot Q_{xy}^2
+\frac{4}{15}\ddot Q_{xz}^2+\frac{4}{15}\ddot Q_{yz}^2\\
r^2\langle{h^2}_{\times}\rangle&=&\frac{1}{6}(\ddot Q_{xx}-\ddot Q_{yy})^2+\frac{2}{3}\ddot Q_{xy}^2
+\frac{4}{3}\ddot Q_{xz}^2+\frac{4}{3}\ddot Q_{yz}^2.
\end{eqnarray}
and
$\langle{\tilde h}_{+}\rangle$ and $\langle{\tilde h}_{\times}\rangle$ are their Fourier transforms.\\
A simpler approach needs to be chosen for the fifth quadrupole derivative in the radiation reaction force. We assume the Quadrupole to be constant in the binary's corotation frame. Therefore the only time dependence arises from the rotation of the coordinate frames which is easy to evaluate and differentiate analytically. This leads to fairly good results as long as we have a binary with two distinct companion stars. However, for a typical merger event, this approximation has shown to underestimate the angular momentum loss by $\simeq40\%$ during GW luminosity peak (see Fig. \ref{fig:jloss2}). Blanchet et al. \cite{BDS} proposed a radiation reaction scheme involving only third order derivatives. However, the quadrupole derivatives and the radiation reaction force formula are only approximate to Newtonian order which rules out at least a direct implementation of this scheme.\\

\begin{figure}[h]
\includegraphics{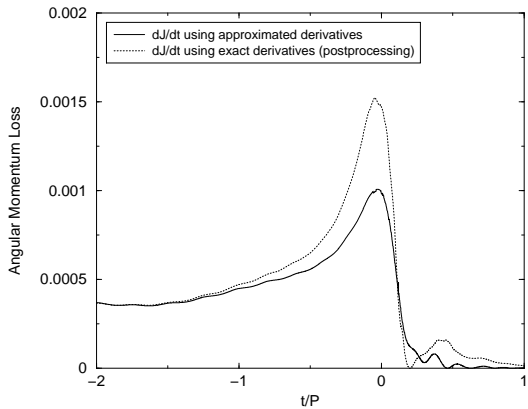}
\caption{Comparison of the angular momentum loss with approximated derivatives and with exact numerical derivatives by postprocessing of the quadrupoles. Shown for model A (see section \ref{sec:merger})}\label{fig:jloss2} 
\end{figure}

\section{Code Calibration}
\label{sec:tests}

In this section, we calibrate our code by examining three important test cases for which data from different approaches is available. The first one is a single rotating polytrope, the second test case is a corotating quasi-equilibrium binary system and the last one tests the gravitational wave luminosity of binary systems.

\subsection{Rapidly Rotating Polytrope}

Axisymmetric configurations in general relativity have been treated in full GR as well as with a CFC approximation scheme \cite{cookGR,cook}. We compare the results obtained against those from \cite{cook}.
To achieve this, we prepare a $\Gamma=3, \bar{M}_0=0.1484$ - neutron star with an density profile corresponding to a static spherical star and an initial velocity field
\begin{equation}
\vec{v}=\vec{\Omega}\times\vec{r}.
\end{equation}
The star is then relaxed into an uniformly rotating equilibrium state with given angular velocity $\Omega$ using the relaxation scheme described in appendix \ref{app:relax}. The values for $\Omega$ have been chosen to fit the values in table I of \cite{cook}.\\
We measure the stars' angular momentum, the gravitational mass, the central energy density $e_c=\rho(1+\epsilon)|_c$ and the ratio of rotational to gravitational energy $T/W$.

\begin{table}[h]
\caption{For each value of the angular velocity $\bar{\Omega}$, we list in the upper line the central energy density $\bar{e}_c$, the gravitational mass $\bar{M}_G$, the angular momentum $\bar{J}_z$ and the ratio of rotational over gravitational energy $T/W$. In the lower line, we list the values from \cite{cook}}.

\begin{ruledtabular}
\begin{tabular}{ccccc}

$\bar{\Omega}$&$\bar{e}_c$&$\bar{M}_G$&$\bar{J}_z$&$T/W$\\
\hline
0.0000&0.960&0.1233&0.0&0.0\\
0.0000&1.000&0.1232&0.0&0.0\\
\hline
0.3315&0.910&0.1236&0.00353&0.0149\\
0.3332&0.903&0.1237&0.00356&0.0149\\
\hline
0.4872&0.824&0.1245&0.00556&0.0355\\
0.4881&0.815&0.1245&0.00559&0.0358\\
\hline
0.5920&0.751&0.1254&0.00726&0.0590\\
0.5927&0.736&0.1254&0.00730&0.0597\\
\hline
0.6630&0.674&0.1265&0.00891&0.0855\\
0.6632&0.665&0.1265&0.00891&0.0858\\
\hline
0.7061&0.610&0.1275&0.01043&0.1122\\
0.7064&0.600&0.1276&0.01049&0.1132\\
\hline
0.7250&0.558&0.1285&0.01178&0.1358\\
0.7256&0.542&0.1287&0.01208&0.1413\\
\hline
\end{tabular}
\end{ruledtabular}
\end{table}

We find generally an agreement within $\sim0.1\%$ in $\bar{M}_G$, $\sim1\%$ in $\bar{J}_z$ and $\sim1\%$ in $T/W$. The more sensitive local quantity $\bar e_c$ agrees to $\sim3\%$.\\
In the last example we faced problems to relax the configuration because the chosen angular velocity is almost at the mass-shedding limit.\\
We used a model with 9939 particles and a (65$\times$65$\times$65) grid. The grid spacing has been chosen to cover the star by $\sim$ 20 mesh cells.

\subsection{Corotating Quasi-Equilibrium Binary systems}
\label{sec:equilibrium}

As a second test case, we have chosen to investigate the properties of corotating quasi-equilibrium binary systems, a system which lies already quite close to our final merger application. We compare our results to \cite{baumgarte}.\\
To construct the system, we set up a pair of Oppenheimer Volkoff stars of equal mass on a given orbit with a fixed binary separation and a guess angular velocity. The relaxation scheme described in appendix \ref{app:relax} drives the system towards a corotating quasi-equilibrium binary system. We store angular velocity, angular momentum, gravitational mass and the relative separation $z=r_{in}/r_{out}$, i.e the ratio of the innermost to the outermost point of the stars.\\
Four relaxation processes with varying binary separation have been performed. The results are listed in table \ref{tab:corot} together with results from \cite{baumgarte}.\\
We find a relative discrepancy of $\sim 0.1\%$ for $\bar{M}_G$, $\sim 0.5\%$ for $\bar{J}_z$ and $\sim 4\%$ for $z$. Again, $z$ as a local quantity is more sensitive. The binary has been modeled with 2$\times$25173 particles and we used a (65$\times$65$\times$65) grid such that one star has been covered by $\sim$ 18 grid cells.

\begin{table}[h]
\caption{For each value of $\bar{\Omega}$, we list in the upper line the relative separation $z$, half of the gravitational mass $\bar{M}_G$ and half of the angular momentum $\bar{J}_z$. In the lower line, we list interpolated values from \cite{baumgarte}.}
\label{tab:corot}
\begin{ruledtabular}
\begin{tabular}{cccc}

$\bar{\Omega}$&$z$&$\bar{M}_G$&$\bar{J}_z$\\
\hline
0.116&0.202&0.14090&0.0418\\
0.116&0.202&0.14086&0.0418\\
\hline
0.123&0.173&0.14090&0.0418\\
0.123&0.169&0.14085&0.0417\\
\hline
0.130&0.135&0.14090&0.0419\\
0.130&0.135&0.14084&0.0416\\
\hline
0.134&0.120&0.14092&0.0419\\
0.134&0.115&0.14084&0.0416\\
\end{tabular}
\end{ruledtabular}
\end{table}

\subsection{Gravitational Wave Luminosity from Quasi-Equilibrium systems}
\label{sec:lgwtest}

The radiation reaction and GW extraction mechanism in our code is independent from the CFC implementation and needs to be calibrated separately. To do so we compare our slow-motion quadrupole formalism (\ref{qpole},\ref{eqn:eloss}) to the results of \cite{duezlgw} which relies basically on a quasi-equilibrium approach in the CFC approximation but with a more accurate GW extraction scheme at the grid boundary.\\
First, we construct an equilibrium configuration using the same method as above. Then, the radiation reaction scheme is switched on and the GW-luminosity is measured. Using the same binary system that was used in \cite{duezlgw}, we calculated four values in a range which was accessible both the QE approach and our code. The results are plotted in Fig. \ref{fig:elosscomp}.

\begin{figure}[h]
\includegraphics{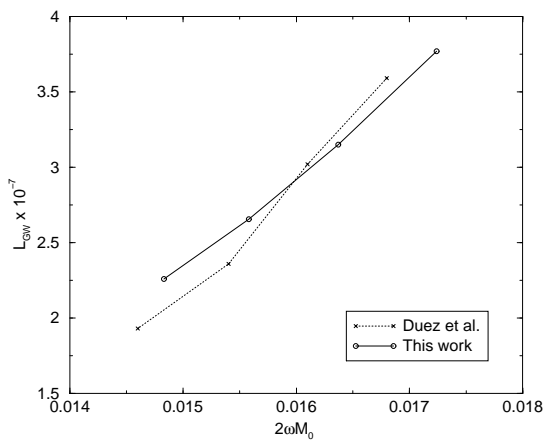}
\caption{Gravitational wave luminosity of a quasi-equilibrium system with $\bar{M}_0$=0.1 and $\Gamma$=2. Duez et al. \cite{duezlgw} extract the GW signal at the boundary of the computational grid while in this work the signal is extracted using the slow-motion quadrupole formula.}\label{fig:elosscomp} 
\end{figure}

Moreover, we test internal consistency by comparing the angular momentum loss due to backreaction force (\ref{radreaction}) and angular momentum loss predicted by the quadrupole formula (\ref{eqn:jloss}). Both curves use the approximated quadrupole derivatives. Fig. \ref{fig:internal} shows that ansatz (\ref{radreaction}) is able to reproduce the slow-motion quadrupole formalism.\\

\begin{figure}[h]
\includegraphics{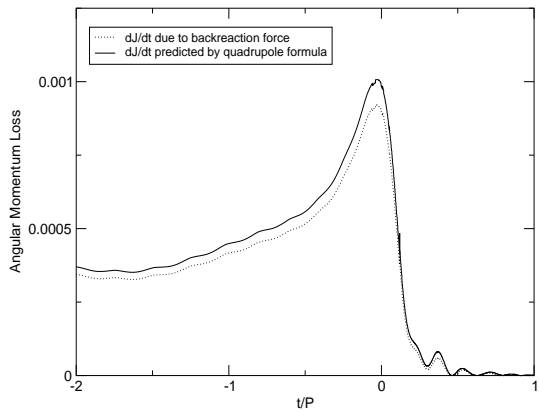}
\caption{Internal consistency test of the backreaction scheme. Compared are the angular momentum loss due to backreaction and the prediction of the quadrupole formula (\ref{eqn:jloss}). Both calculations use the approximated quadrupole derivatives.}\label{fig:internal} 
\end{figure}

\section{Binary Neutron Star Merger}
\label{sec:merger}

Having done the tests, we apply the code to the merger of two NS in a binary system which are modeled as simple polytropes. Three models are considered with a different adiabatic index each time.
We analyze morphology, gravitational wave signal, gravitational luminosity spectrum and amount of ejected material depending on the stiffness of the EoS.

\subsection{Initial Setup}

For a model with a given adiabatic index $\Gamma$, we have basically two free parameters to choose, the rest mass $M_0$ and, if we want to work in physical quantities, the polytropic constant $\kappa$. We assign to the models with a stiff EoS a larger rest mass since as a stiff EoS allows a much larger maximal mass than a soft one. The polytropic constant $\kappa$ is chosen to match the radii of the three models to $R_0=9.6$km.
To choose a optimal initial binary separation, the ISCO has been determined experimentally by relaxing the binary systems in the mutual field. By varying the binary separation is steps of $0.5$km, we find the closest binary which is still stable against tidal disruption but whose separation is close to the ISCO to save computational resources.\\
For all three models, we use 50346 SPH particles in total for the hydrodynamics and a ($65\times 65\times 65$) grid to solve the field equations. As the central high density region forms, we switch to a nested grid configuration (see Fig. \ref{fig:grids}). The relevant parameters are summarized in table \ref{tab:parameters}.

\begin{table}[h]
\caption{Input parameters for models A - C. Listed are the adiabatic index $\Gamma$, polytropic constant $\kappa$, the ratio of binary separation at the start of simulation to the radius of a spherical star $d_{initial}/R_0$, the rest mass of one star, half of the total gravitational mass, the ratio of rest mass of one star to the maximal mass of the OV star of the corresponding model $C_{mass}=M_0/M_0^{max}$}
\label{tab:parameters}
\begin{ruledtabular}
\begin{tabular}{ccccccc}
Model & $\Gamma$ & $\kappa$ & $d_{initial}/R_0$ & $M_0$ & $M_G/2$ & $C_{mass}$\\
\hline
A & 2.0 & $7.032\times10^1$ & 3.02 & 1.34 & 1.235 & 0.89\\
B & 2.6 & $5.600\times10^3$ & 3.13 & 1.60 & 1.421 & 0.70\\
C & 3.0 & $1.071\times10^5$ & 3.47 & 1.74 & 1.555 & 0.65\\
\end{tabular}
\end{ruledtabular}
\end{table}

\begin{figure}[h]
\includegraphics{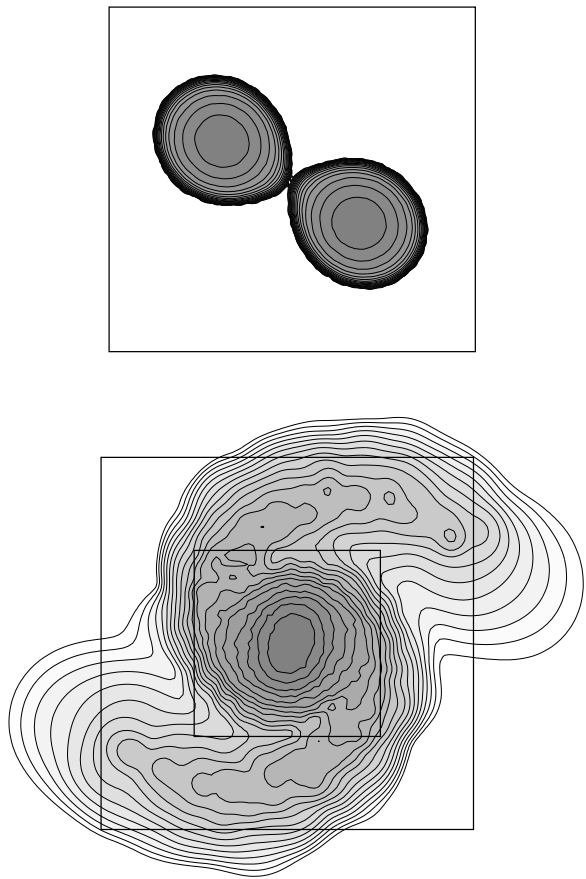}
\caption{Gravity grid setup during the simulation. The top figure sketches the single-grid setup with a ($65\times 65\times 65$) grid during the pre-merger phase. The bottom figure shows the two-level-nested grid configuration with two ($65\times 65\times 65$) grids. }\label{fig:grids} 
\end{figure}

In the following discussion we measure distance in units of $R_0$, density in units of $\rho_0$, the central density of the OV profile, time in units of P, the rotation period of the binary where $d=3R_0$, mass and energy in units of $M_0$ and the GW-frequency is measured in units of $f_0=2/P$. Moreover, we choose the origin of our time axis at the GW luminosity peak. To allow the reader a conversion into physical units, we summarize the values of the normalization constants in table \ref{tab:normconsts}

\begin{table}[h]
\caption{Normalization constants used in the discussion}
\vspace{0.2cm}
\label{tab:normconsts}
\begin{ruledtabular}
\begin{tabular}{cccccc}
Model & $R_0$ [km] & $\rho_0$ [gcm${}^{-3}$] & P [ms] & $M_0$ [M${}_\odot$] & $f_0$ [Hz]\\
\hline
A&9.6&1.30$\times10^{15}$&2.051&1.34&975.27\\
B&9.6&8.66$\times10^{14}$&1.955&1.60&1023.27\\
C&9.6&7.64$\times10^{14}$&1.935&1.74&1033.58\\
\end{tabular}
\end{ruledtabular}
\end{table}

\subsection{Morphology}
\label{sec:morphology}
The simulation is started from the quasi-equilibrium configuration outside the ISCO. Since this setup contains no inward radial velocity, the binary needs initially about half a revolution to start with the orbital decay. The stars then slowly approach each other and merge within about three orbital revolutions.\\
Shortly after the two stellar cores begin to merge, the gravitational wave luminosity reaches its maximum, followed by the minimum as the two cores begin to build up the central object. At the same time, the system starts to shed mass through the outer Lagrangian points and builds up spiral arms around the newly forming central object. Contrary to Newtonian simulations \cite{max,rossi2001}, the spiral arms do not extend far out into space, but get closely wrapped around and finally sink back into the disk of matter around the merger site. In the post-merger phase, all models lead to a differentially rotating central object which first bounces and then slowly contracts (see Fig. \ref{fig:rhoc}). The deviation from axisymmetry disappears within 3 (model A) to 7 (model C) revolutions.

\begin{figure}[h]
\includegraphics{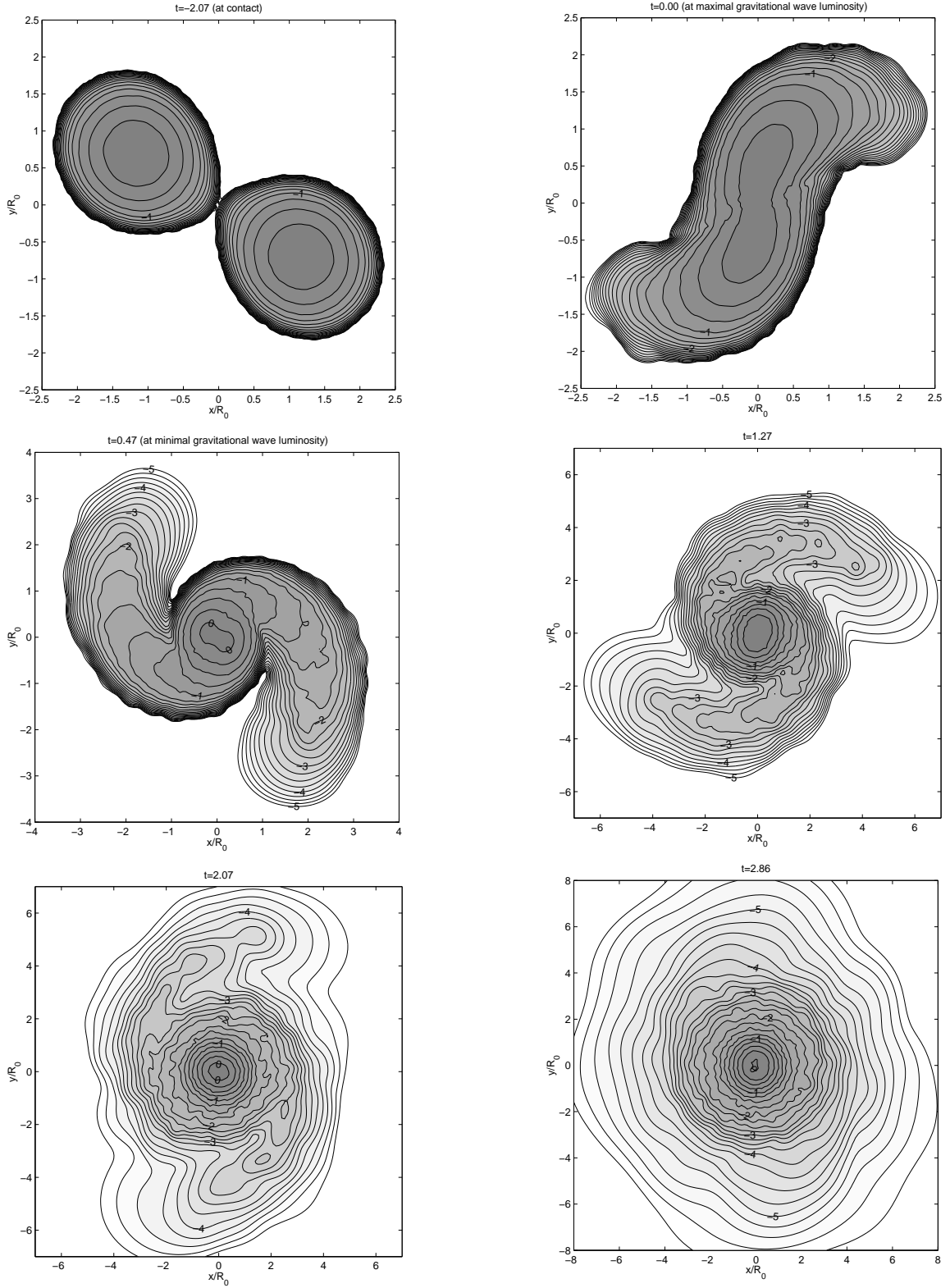}
\caption{Normalized density distribution in the orbital plane of Model A. The contour lines range from $10^{-5}$ to $10^{0}$ with a logarithmic spacing of 0.25 dex.}\label{fig:morph20}
\end{figure}

\clearpage

\subsection{Evolution of Angular Momentum and Gravitational Mass}

An important test of every hydrodynamical code is the check for conserved quantities. In our case, we store angular momentum and gravitational mass which would be conserved to 1PN order in the absence of backreaction forces (\ref{radreaction}). In Fig. \ref{fig:angmom} we plot the two quantities versus time for the three models. Likewise, we plot the same quantities, but with the integrated angular momentum and energy loss (\ref{eqn:jloss},\ref{eqn:eloss}) subtracted to follow the net conservation without backreaction.\\
Concerning angular momentum, we see in all models a loss of $\sim 8\%$ arising from the backreaction force. In the post-merger stage, this force is artificially suppressed in our calculations due to our approximation of the quadrupole derivatives (see Fig. \ref{fig:jloss2}) and the loss in the angular momentum of $\sim 0.5\%$ is mainly due to numerical errors caused by the limited resolution of the gravity grid. The corrected angular momenta are conserved within $\sim 1\%$ during the whole merger process. This is obviously well within the expected 1PN accuracy range of $(M/R)^2\sim$ 5$\times 10^{-2}$.\\
A similar behaviour is observed for the gravitational masses. After a decrease of $\sim 0.4\%$ due to GW radiation, the masses of all models decrease by $\sim 0.2\%$ due to the same numerical errors. The small jumps are due to the resizing of the gravity grid during the run as the expression for $M_G$ contains the non-compact term $K_{ij}K^{ij}$. The corrected masses vary within $\sim 0.2-0.3\%$ over the whole process.
\begin{figure}[h]
\includegraphics{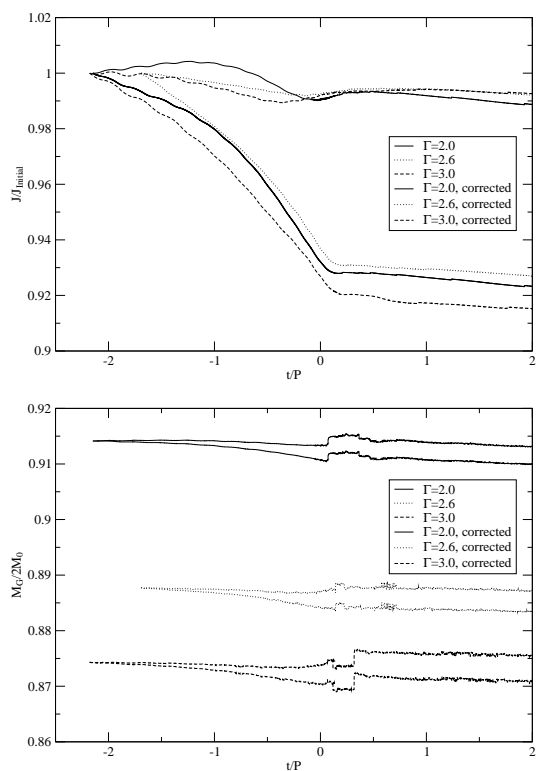}
\caption{Plotted is the temporal evolution of the total angular momentum and gravitational mass. The angular momentum is normalized to the initial value, the gravitational mass to total rest mass. Also indicated are the same corrected quantities but with the integrated angular momentum and energy loss subtracted to show the net conservation. The angular momentum decrease due to gravitational radiation is $\sim 8\%$ but the corrected angular momenta are conserved within $\sim 1\%$. The jumps in the gravitational mass occur due to resizing of the gravity grid setup.}\label{fig:angmom}
\end{figure}


\subsection{Maximal densities and Black hole formation}

For corotating quasi-equilibrium sequences, a decrease of the maximal density with decreasing binary distance is predicted \cite{baumgarte,wilsonshort}. We therefore spend special care to the maximal density evolution during inspiral and relax the binary as carefully as possible prior to the simulation. In Fig. \ref{fig:rhoc}, we plot $\rho_{max}$ versus time. During the inspiral phase, we see for all models a nearly constant evolution with a small increase of $\sim 1\%$ for model C, followed by the sharp decline at $t/P\simeq0$ as the stars get disrupted. This effect is comparable to the numerical accuracy of the code.\\
At the end of the merger phase, a dense central object builds up. It contracts but soon bounces back due to pressure and centrifugal forces. This bounce can be seen in all three models by a peak in central density, i.e. none of the models does immediately collapse to a BH. Afterwards, an oscillating central object with rising central density forms. However the grid resolution to calculate the gravity during the ringdown stage is not sufficient. Test calculations with enhanced resolution have shown increased self-gravity forces and higher central densities. We therefore consider the central densities in the present simulation as a lower limit.\\

\begin{figure}[h]
\includegraphics{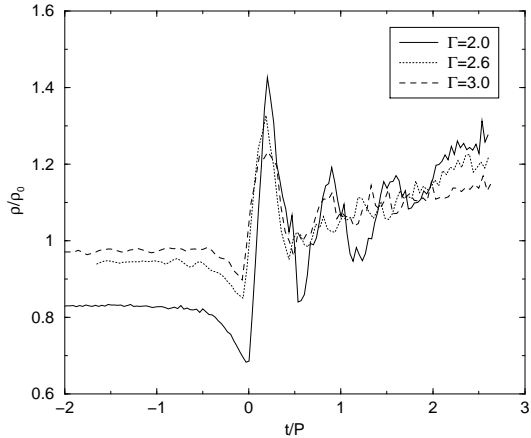}
\caption{Time evolution of the maximal density in the three models. The values are normalized to the central densities $\rho_0$ of the OV star corresponding to the model. The soft model starts with the lowest maximal density of the three since the tidal distortion is largest in this case. The maximum drops to a minimum as the cores begin to merge. Also visible in all models is a central bounce just after the merger phase. In the ringdown phase of the central object, the maximal density steadily rises without convergence, towards a possible later collapse.}\label{fig:rhoc}
\end{figure}

We can also consider the dimensionless rotation parameter $a=J_z/M_G^2$ of the system. If this parameter is less that unity, in most cases, except for very stiff EoSs, a black hole will be formed. This parameter drops below unity for all three models after the merger phase.\\
Model A can be compared to model (C2) of \cite{shibatauryu}. The parameters $a_{initial}$, the rotation parameter at the beginning of the simulation, and $C_{mass}$, the ratio of the rest mass of one star to the maximal mass for a spherical star allowed by the EoS agree within $\simeq 2\%$ and we therefore expect a qualitative agreement. However, \cite{shibatauryu} see a collapse to a BH on a dynamical timescale without a bounce contrary to this work. We see three possible reasons for the different outcome.
\begin{itemize}
\item The numerical resolution is comparable in both models, but not before we switch to the nested grid mode. This happens just before the bounce, at about $t/P\simeq0.1$. Therefore, it is well possible, that we underestimate the central densities in the moment when an eventual collapse builds up.
\item The slightly different dynamical evolution during the merger phase when the CFC approximation deviates most from full GR could have a direct influence.
\item Since differential rotation plays an essential role in the stability of supramassive stars, we investigated the rotational pattern of the central object. We fit the angular velocities $\omega_i$ to a typical $\omega(r)=\omega_c/(1+r_{cyl}^2A^{-2})$ - law \cite{baumgartediffrot} to get a quantitative estimate and obtain a value for $A$ which is a measure for the amount of differential rotation. For model A, the result is $A\simeq1.7R_0$, i.e $\hat{A}=A/R_e\simeq1.1$ if we take $R_e\simeq1.5R_0$ for the equatorial radius of the central object. But an object with a value of $\hat{A}^{-1}\simeq0.9$ could well support a rest mass larger than $3M_{\odot}$\cite{baumgartediffrot}. This is a strong argument against the collapse in our calculations, a difference in the numerical viscosity might therefore explain the different outcome.
\end{itemize}
\begin{figure}[h]
\includegraphics{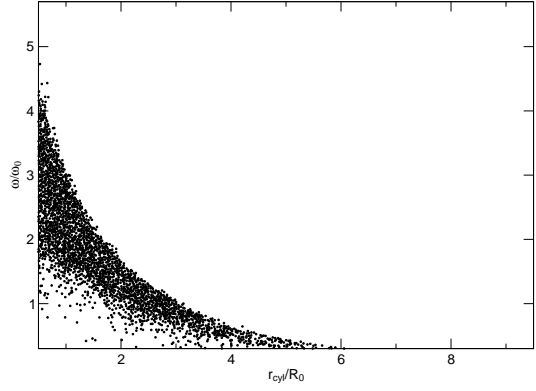}
\caption{Plotted are the angular velocities $\omega_i$ vs. the cylindrical radius at $t/P=1.19$ for model A. The angular velocity is normalized to the angular velocity where $d=3R_0$}\label{fig:diffrot}
\end{figure}

\subsection{Ejected Matter}

With respect to NS mergers being a promising site for r-process, a crucial question is the amount of matter that becomes unbound and gets ejected into space. In Newtonian simulations, a small fraction at the tip of the spiral arms has enough total energy to leave the central gravitational potential. To determine the amount of unbound matter in the relativistic case, we need a criterion similar to the Newtonian energy-criterion. To do so, we assume the following conditions for a spiral arm particle: 
\begin{itemize}
\item Pressure forces are negligible with respect to gravitational forces (i.e. the particles move on geodesics)
\item Particles move in a stationary metric (which is produced mainly by the axisymmetric, rapidly rotating central object)
\end{itemize}
Then, we can find a conserved expression $\epsilon_{stationary}$ which is similar to the Newtonian $e_{kin}+e_{int}+e_{pot}$-Expression for total energy per unit mass
\begin{equation}
\epsilon_{stationary}=v^i\tilde{u}_i+\frac{\epsilon}{u^0}+\frac{1}{u^0}-1
\end{equation}
The derivation can be sketched as follows. Using the momentum eqn. (\ref{momentumeqn}) we get
\begin{equation}
\rst\frac{d}{dt}(v^i\tilde{u_i})=\rst\frac{d}{dt}(\beta^i\tilde{u_i})-\frac{d}{dt}({\tilde{u^0}})\rst\frac{\tilde{u_k}\tilde{u_k}}{(\tilde{u^0})^2\psi^4}-2\alpha\rst\tilde{u^0}\dot{\alpha}
\end{equation}
where we used $\partial_t\alpha\equiv0$, $\partial_t\beta^i\equiv0$, $\partial_t\psi\equiv0$, $p\equiv0$ and the time derivative of the identity of $v^i=-\beta^i+\frac{u_i}{\psi^4 u^0}$. On the other hand with the aid of the energy eqn. (\ref{energyeqn}) we get
\begin{equation}
2\rst\frac{d}{dt}(\frac{\epsilon}{u^0})=-2\rst\frac{\epsilon\dot u^0}{(u^0)^2}.
\end{equation}
Adding up the two identities and using eqn. (\ref{norm}) we arrive at
\begin{equation}
\rst\frac{d}{dt}(v^i\tilde{u_i}-\beta^i\tilde{u_i}+2\frac{1}{u^0}+\frac{w}{u^0}\frac{u_iu_i}{\psi^4}+2\frac{\epsilon}{u^0})=0
\end{equation}
which simplifies to
\begin{equation}
\rst\frac{d}{dt}(v^i\tilde{u_i}+\frac{\epsilon}{u^0}+\frac{1}{u^0})=0.
\end{equation}
To estimate how much mass will be ejected, we plot $\epsilon_{stationary}$ as a function of the distance to the central object at the end of the simulation. There are only several distinct particles having positive energy corresponding to a cumulative rest mass of approximative 1.5$\times$10$^{-4}$M${}_\odot$ (9 particles), 5$\times$10$^{-5}$M${}_\odot$ (2) and 2$\times$10$^{-4}$M${}_\odot$ (5) for models A, B and C, respectively.\\
These values can only be taken as an order of estimate since the resolution in the spiral arm region is far too low. Compared to model C in \cite{rosswog99} ($M_{ej}\simeq$1.5$\times$10$^{-2}$M${}_\odot$) which is the Newtonian analogon to model B of this work, the relativistic effect of a stronger gravitational attraction clearly becomes apparent.\\
It has been stated, that given the uncertainties in the merger rates, even an amount as small as a few times 10$^{-4}$M${}_\odot$ per event in r-process material may be an important contribution to the enrichment of the Galaxy with heavy elements (see Fig. 26, in \cite{rosswog99}, with an upper limit to the merger rate of 10$^{-4}$ per year and galaxy \cite{kalogera}). Therefore, further, high-resolution calculations are needed for precise answers. 

\begin{figure}[h]
\includegraphics{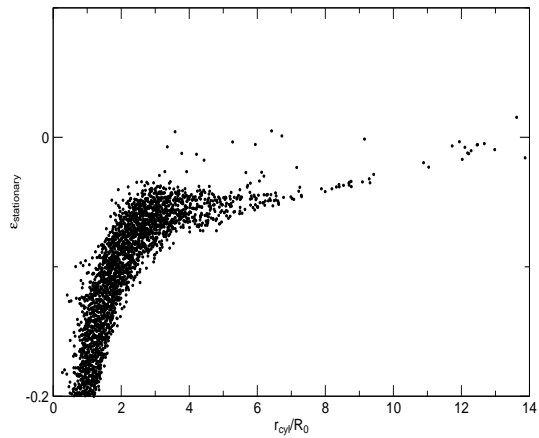}
\caption{Plotted are the values of $\epsilon_{stationary}$ versus the cylindrical radius $r_{cyl}=\sqrt{x^2+y^2}$ at $t/P=1.19$ for model A. The particles with values below zero are considered to be bound. Only very few particles are unbound corresponding to some 10${}^{-4}$M${}_{\odot}$. The situation is similar for model B and C.}
\end{figure}

\subsection{Gravitational Radiation}

The gravitational radiation waveform as well as the gravitational wave frequency spectrum are expected to contain important information about the dynamics of the merger and the internal structure of the NS, especially the EoS. In the following, we compare the radiation waveform and frequency spectrum depending on the stiffness of the EoS.\\
The gravitational radiation waveforms are shown in Fig. \ref{fig:gwaves}. To allow for comparison, we matched the simulated waveform onto a pointmass result which is based basically on Newtonian gravity \cite{MTW}, but with the inspiral time $\tau$ fitted to the code signal and with the quadrupoles (\ref{qpole}) used. Apart from a small accumulating phaseshift which results from the Newtonian approximation, we notice that the code signals follow remarkable closely the pointmass signal until only a few oscillations before merging. Partly a reason for this effect is the fact that we work in the slow motion limit. Only the $\Gamma=2$ signal shows an exceptional behaviour as it exceeds the pointmass signal in amplitude and frequency at about $t/P\simeq-0.4$, two oscillations before merging. This is caused by a dramatical acceleration of the orbital decay and a speedup of the binary in angular velocity. We do not see the effect in such a drastic form in the stiffer models B and C. The reason for the special behaviour of model A is a numerical one: the error in angular momentum is up to one third of the backreaction loss during the last orbit which causes the observed speedup. A comparative run to model A with a slightly different numerical setup shows a different angular momentum violation behaviour. This results in a different acceleration of the orbital decay and the angular velocity and in slightly different wave signals.\\
A clear difference between the three models, which has also been pointed out in the Newtonian case \cite{RS2}, is the intensity and length of the ringdown signal. The GW signal of the soft-model A shows a distinct peak before decaying rapidly within $\simeq$10 oscillations, whereas the stiffer models are characterized by larger peaks and a decay on a longer timescale which exceeds the simulation timescale in model C. Contrary to other investigations \cite{shai, faber}, a third peak cannot be seen.\\
All these phenomena can also be observed in the luminosity spectrum Fig. \ref{fig:dldf}. We indicate the GW-frequencies at dynamical instability and at maximal gravitational wave luminosity, $f_{dyn}$ and $f_{GW}$, respectively. The quadrupole oscillation frequency of final central object is denoted by $f_{Qpole}$. In the frequency range below $f_{dyn}$, the three models follow the pointmass behaviour $dE/df(f)\simeq f^{-1/3}$. Beyond this frequency, model A begins to deviate, due to the same numerical reason mentioned above: while the angular velocity increases, the merging process is accelerated and less energy (which is the time integral of the GW-luminosity) will be accumulated in one frequency interval. The other two models follow the pointmass template up to about $f_{GW}$. The spectrum of the comparative run to model A also deviates from the pointmass spectrum at $f_{dyn}$ but lies above that of model A. At $\log(f/f_0)\simeq0.45$, the curve then bends down to the minimum.\\
Since all our simulations underestimate the backreaction force in the late merger phase, the real merging process might exhibit a larger acceleration than the one in our simulations. Therefore, we expect that the real spectra will lie below ours and therefore also below the pointmass spectra in the frequency range between $f_{dyn}$ and $f_{GW}$. Such a behaviour has been observed in Newtonian simulations \cite{zhuge2}.\\
The ringdown signals translate into strong maxima, peaked around the quadrupole oscillation frequency $f_{Qpole}$. Here we observe clear differences in strength between the three models. While the stiff-EoS model C peaks well above the pointmass line the peak of model A is much weaker. This has already been shown in Newtonian \cite{zhuge2} and Post-Newtonian \cite{faberconf} investigations. Moreover the normalized quadrupole frequency $f_{Qpole}/f_{0}$ in model A is slightly higher than in the other two models since in model A, the soft EoS produces a much more centrally condensed central object.\\
Therefore, we are able to distinguish the stiff-EoS spectrum from the soft-EoS spectrum by means of the ringdown part. However, as long as the numerical inaccuracies during the merger phase are still of the order of the backreaction terms, the exact shape of the part below $f_{GW}$ cannot be reliably determined.\\
Contrary to other investigations in PN approximation \cite{faberconf}, we do no observe a dip between $f_{dyn}$ and $f_{GW}$. The difference may be explained in the different pointmass waveform fitting procedure. While \cite{faberconf} adds a purely Newtonian waveform, we modify the inspiral time $\tau$ and the quadrupoles. This increases the frequency of the pointmass waveform so the dip is filled up.

\begin{figure}[h]
\includegraphics{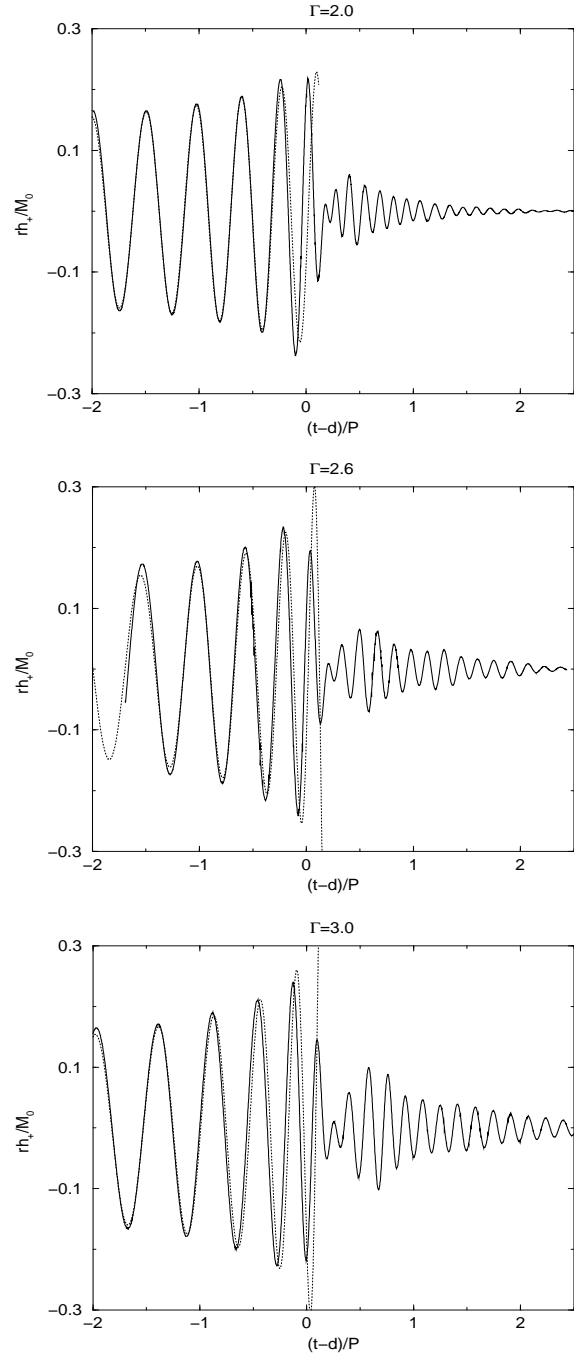}
\caption{Gravitational waveforms of models A to C normalized to the rest mass. The time axis is normalized to the orbital period $P$ at a binary distance $d=3R_0$. The dashed lines correspond to fitted pointmass waveforms.}\label{fig:gwaves}
\end{figure}

\begin{figure}[h]
\includegraphics{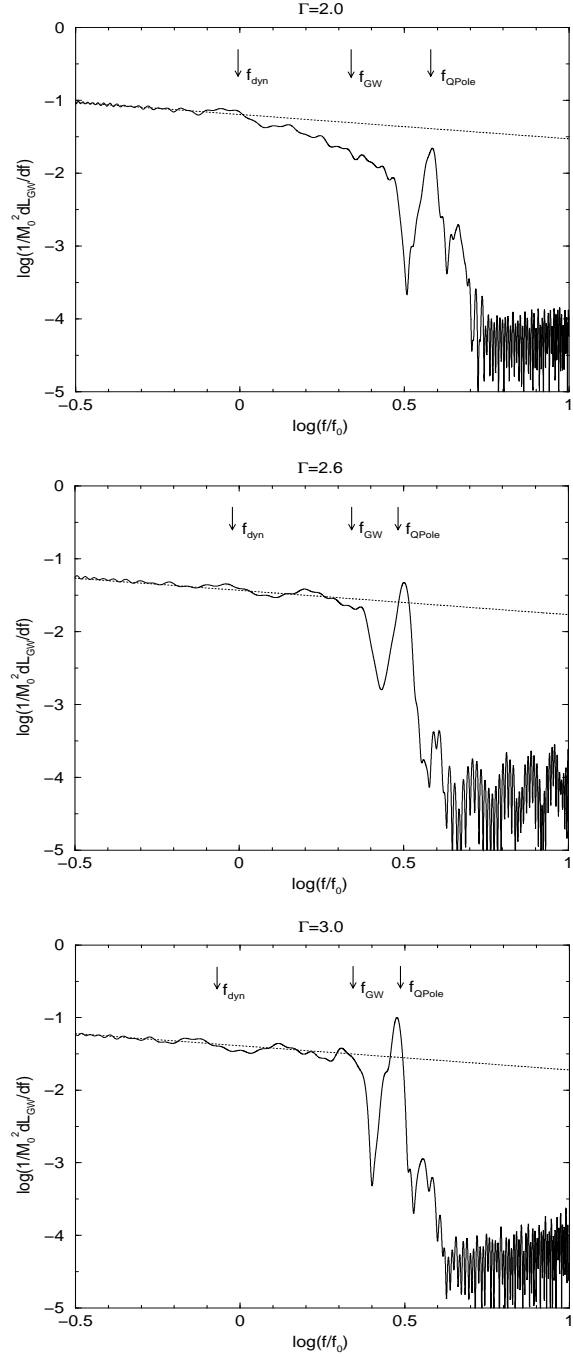}
\caption{Frequency spectra for the three different models. The dashed line shows the pointmass spectrum $dE/df\simeq f^{-1/3}$. The frequency axis is normalized to the GW frequency $f_0$ at a binary distance $d=3R_0$. Model A deviates early from the pointmass behaviour because of numerical errors. Due to a build-up of a non-axisymmetric central object, the stiff-EoS model C produces a much higher quadrupole oscillation peak than the soft-EoS model A, whose central object quickly relaxes to axisymmetry. Furthermore, as model A produces a more centrally condensed object, the quadrupole oscillation peak is shifted to slightly higher frequencies.}\label{fig:dldf}
\end{figure}

\clearpage

\begin{figure}[h]
\includegraphics{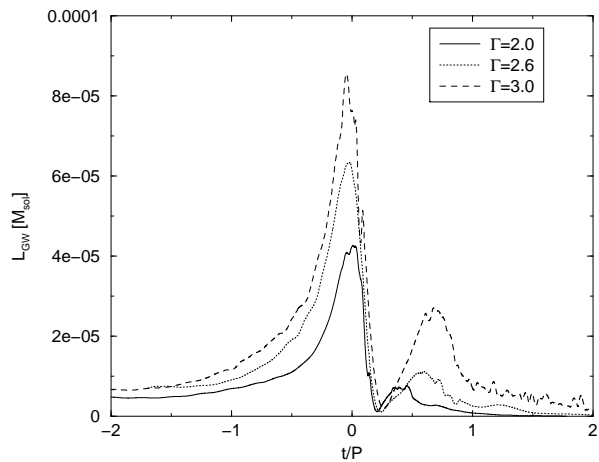}
\caption{Gravitational wave luminosities for the three models considered. All models show a second peak but no further peaks can be detected. The strength increases with higher stiffness.}\label{fig:eloss}
\end{figure}

\section{Conclusion}
\label{sec:conclusion}

We have developed and tested a 3D smoothed particle hydrodynamic code based on the conformally flat approximation to GR with an additional approximative backreaction scheme. The gravity field equations are solved on an overlaid grid with a multigrid solver. The code has been applied to the neutron star merger problem using a polytropic equation of state. Three models have been considered, with adiabatic indices $\Gamma=2.0$, $\Gamma=2.6$ and $\Gamma=3.0$. We start with quasi-equilibrium models to avoid oscillations. In all models, we find about 8\% of angular momentum being radiated away due to backreaction.\\
None of the central objects does immediately collapse to a black hole. Instead, we always find a bounce followed by a slow contraction. We also find a highly differentially rotating object in all three models.\\
The amount of unbound mass is significantly less than obtained in Newtonian simulations. We obtain values up to some $10^{-4}$M${}_\odot$. This mass corresponds to only a small number of particles, therefore final conclusions cannot be drawn.\\   
We are able to distinguish the gravitational wave signal and spectrum of a stiff-EoS model from a soft-EoS model by means of the ringdown part because the stiff-EoS model produces a much stronger signal during this phase. In general, however, we are not yet able to determine the exact shape of the spectrum due to numerical errors caused by the limited resolution of the overlaid field solving grid and a too simple evaluation of the quadrupole derivatives around the gravitational radiation peak. 
In future work we will increase the resolution by increasing the particle number and using a finer field solving grid to make the results more reliable. The code shall also be generalized to nonsynchronized binaries, an initial condition which is likely the case for real systems. Moreover, we plan to incorporate a realistic EoS (e.g. \cite{shen}).

\begin{acknowledgments}
RO and FKT are funded by the Schweizerischer Nationalfonds under grant 2000-061822.00,\\
SRO is funded by a PPARC rolling grant for Theoretical Astrophysics.
\end{acknowledgments}

\appendix

\section{Conformally Flat Formalism}
\label{app:formalism}
We shortly summarize in this section the formalism on which our code is based. It follows closely the derivation in \cite{shibatacfc} and \cite{baumgarte}, but is slightly adapted to a Lagrangian form suitable for SPH. Concerning indices, we use the summation convention where Latin indices run from 1 to 3 and Greek indices from 0 to 3. We also set $G=c=1$. 

\subsection{Conformally flat approximation}

The spatial part $\gamma_{ij}$ of the metric is approximated by the conformally flat form
\begin{equation}
\gamma_{ij}=\psi^4\delta_{ij}
\end{equation}
so that the ADM line element reads
\begin{equation}
ds^2=(-\alpha^2+\beta_i\beta^i)dt^2+2\beta_i dx^i dt+\psi^4\delta_{ij}dx^i dx^j.
\end{equation}
Here, $\alpha$ is the lapse function, $\beta_i$ the shift vector and $\psi$ the conformal factor.

\subsection{Hydrodynamic equations}

The fluid is described with a perfect fluid stress-energy tensor
\begin{equation}
T_{\mu\nu}=\rho wu^\mu u^\nu+p g^{\mu\nu}
\end{equation}
where $\rho$ refers to the rest mass density, $w=1+p/\rho+\epsilon$ to the specific enthalpy, $p$ to the pressure, $\epsilon$ to the specific internal energy, and $u^\mu$ to the four velocity of the fluid.\\
With the coordinate conserved density
\begin{equation}
\rst=\rho\alpha u^0\psi^6\label{stardens}
\end{equation}
the continuity equation reads
\begin{equation}
\label{conteqn}
\frac{d}{dt}\rst+\rst\pa{i} v^i=0,
\end{equation}
where $v^i=u^i/u^0$ is the physical velocity. The relativistic momentum equation reads
\begin{equation}
\label{momentumeqn}
\frac{d}{dt}\tilde{u}_i=-\frac{1}{\rst}\alpha\psi^6\pa{i} p-\alpha\tilde{u}^0\pa{i}\alpha+\tilde{u}_j\pa{i}\beta^j+\frac{2\tilde{u}_k\tilde{u}_k}{\psi^5\tilde{u^0}}\pa{i}\psi,
\end{equation}
where $\tilde{u}_\mu=wu_\mu$ is the specific momentum which connects to the physical velocity by $v^i=-\beta^i+\frac{u_i}{\psi^4 u^0}$. Moreover, we get from the normalization condition $u_\mu u^\mu =-1$
\begin{equation}
\label{norm}
(\alpha u^0)^2=1+\frac{u_i u_i}{\psi^4}.
\end{equation}
An energy conservation equation enables us to evolve the internal energy
\begin{equation}
\label{energyeqn}
\frac{d}{dt}\epsilon=-\frac{p}{\rho}\pa{i}v^i-\frac{p}{\rho}\frac{d}{dt}\ln(\alpha u^0\psi^6),
\end{equation}
and finally, an equation of state
\begin{equation}
p=p(\rho,\epsilon)
\end{equation}
closes the system.

\subsection{Field equations}

Equations for the metric components $\alpha$, $\beta^i$ and $\psi$ can be found from the maximal slicing assumption, the Hamiltonian constraint and the momentum constraint.\\

The maximal slicing assumption tr$K_{ij}=0$, where $K_{ij}$ is the extrinsic curvature, leads to a Poisson-like equation
\begin{equation}
\Delta(\alpha\psi)=2\pi\alpha\psi^5(E+2T_{ij}\delta_{ij}\psi^{-4})+\frac{7}{8}\alpha\psi^5 K_{ij}K^{ij},
\label{alphapsieqn}
\end{equation}
where
\begin{equation}
E=\rho w(\alpha u^0)^2-p
\end{equation}
and
\begin{equation}
\label{kijform}
2\alpha\psi^{-4}K_{ij}=\partial_j\beta^i+\partial_i\beta^j-\frac{2}{3}\delta_{ij}\partial_k\beta^k.
\end{equation}
The Hamiltonian constraint serves us an equation for the conformal factor
\begin{equation}
\Delta\psi=-2\pi\psi^5 E-\frac{1}{8}\psi^5 K_{ij}K^{ij}\equiv 4\pi S_{\psi}.
\label{psieqn}
\end{equation}
And, finally, with the momentum constraint we obtain an equation for the shift vector
\begin{equation}
\Delta\beta^i+\frac{1}{3}\partial_{ij}\beta^j=\pa{j}\ln\left(\frac{\alpha}{\psi^6}\right)\left(\pa{j}\beta^i+\pa{i}\beta^j-\frac{2}{3}\delta_{ij}\pa{l}\beta^l\right)+16\pi\alpha S_i\equiv 16\pi\alpha S^0_i
\label{momentum}
\end{equation}
where $S_i=\rst\tilde{u}_i\psi^{-6}$.
Using the definition \cite{baumgarte}
\begin{equation}
\beta^i=B^i-\frac{1}{4}\pa{i}\chi,
\end{equation}
(\ref{momentum}) splits into two simpler parts
\begin{equation}
\Delta B^i=\pa{j}\ln\left(\frac{\alpha}{\psi^6}\right)\left(\pa{j}\beta^i+\pa{i}\beta^j-\frac{2}{3}\delta_{ij}\pa{l}\beta^l\right)+16\pi\alpha S_i
\label{beqn}
\end{equation}
and
\begin{equation}
\Delta\chi=\pa{j}B^j.
\label{chieqn}
\end{equation}
Furthermore, we extract the following physical observables
\begin{eqnarray}
M_0&=&\int\rst d^3x,\\
J_i&=&\epsilon_{ijk}\int\rst x^j\tilde{u}_kd^3x,\\
M_G&=&-2\int S_{\psi}d^3x,
\end{eqnarray}
which correspond to the total rest mass, total angular momentum and total gravitational mass. We have to be aware, that these quantities are conserved only to 1PN order, the order of the CFC method. Therefore, measurements of these quantities are only accurate to order $(M/R)$.

\subsection{Boundary conditions}

To solve the equations for $\alpha\psi$ (\ref{alphapsieqn}) and $\psi$ (\ref{psieqn}), we expand the source around both stars in multipoles and approximate the values on the outer boundary of the computational grid up to the hexadecupole order. Note that due to the finite grid size, it is not possible to include the contribution from $K_{ij}K^{ij}$ which extends beyond the grid boundary. However, by comparing the contribution from $K_{ij}K^{ij}$ on different grid sizes, and using its $r^{-6}$-fall-off-dependence, which follows from the asymptotic form of the potentials and (\ref{kijform}), we assume the exterior contribution to be $\lesssim 0.01$ M${}_\odot$.\\
The same argument, on the other hand, does not hold for the shift-vector equations (\ref{beqn}) and (\ref{chieqn}) since the non-compact-support source term $\pa{j}\ln(\alpha\psi^{-6})\left(\pa{j}\beta^i+\pa{i}\beta^j-\frac{2}{3}\delta_{ij}\pa{l}\beta^l\right)$ is not small compared to the matter source term and we therefore cannot calculate exactly the integrals over the source function $S^0_i$. A possible solution is to impose only fall-off boundary conditions \cite{baumgarte} (note: use a different coordinate system) which makes them independent of the absolute values of the exact integrals.
\begin{eqnarray}
B^x&\sim&\frac{x}{r^3}\int_{G}{x \alpha{S}^0_x}d^3x+\frac{y}{r^3}\int_{G}{y \alpha{S}^0_x}d^3x\equiv b^x\label{approxbx}\\
B^y&\sim&\frac{x}{r^3}\int_{G}{x \alpha{S}^0_y}d^3x+\frac{y}{r^3}\int_{G}{y \alpha{S}^0_y}d^3x\equiv b^y\label{approxby}\\
B^z&\sim&\frac{xyz}{r^7}\equiv b^z\label{approxbz}\\
\chi&\sim&\frac{xy}{r^5}\equiv c\label{approxchi}
\end{eqnarray}

\subsection{Backreaction Force and Gravitational Wave Extraction}

Like the Newtonian and the PN approximation, the CFC approximation does not include gravitational radiation, an additional scheme to mimic the loss of energy and angular momentum and an extraction scheme of gravitational waves needs to be implemented. The radiation reaction force is modeled via an ansatz which generalizes the Burke-Thorne radiation reaction formula. The gravitational wave extraction scheme is approximated to quadrupole order.\\
For the radiation reaction force, we propose the following ansatz
\begin{equation}
F_i^{reac}=\sigma\pa{i}V^{reac}
\label{radreaction}
\end{equation}
where
\begin{eqnarray}
V^{reac}&=&-\frac{1}{5}x_ix_j(Q_{ij})^{(5)}
\end{eqnarray}
and $\sigma$ is the active gravitational mass density in the CFC approximation
\begin{equation}
\sigma=T^{00}+T^{ii}=\rho(1+\frac{p}{\rho}+\epsilon)(u^0)^2(1+\mathbf{v}^2)+p(-\frac{1}{\alpha^2}-\frac{\beta^i\beta^i}{\alpha^2}+3\psi^{-4}).
\end{equation}
The quadrupole finally is approximated \cite{wilson}
\begin{equation}
Q_{ij}=\text{STF}\left\{-2\int S_{\psi}(x)x_i x_j d^3x\right\}.
\label{qpole}
\end{equation}
where $S_{\psi}(x)$ is the source term in equation (\ref{psieqn}). We recover the quadrupole in the so-called slow-motion approximation which is valid for slowly moving strong-field sources \cite{thorne80}.
The symbol 'STF' refers to 'symmetric trace-free part' which can be calculated in the case of a two-index quantity as
\begin{equation}
\text{STF}A_{ij}=\frac{1}{2}A_{ij}+\frac{1}{2}A_{ji}-\frac{1}{3}\delta_{ij}A_{kk}.
\end{equation}
This approximation is based on the idea to expand a metric perturbed by radiation into mass and current multipoles of the radiating source \cite{thorne80}.
Using the above quadrupole, we get up to quadrupole order
\begin{eqnarray}
\label{eqn:jloss}
\dot J_i&=&-\frac{2}{5}\epsilon_{ijk}(Q_{jm})^{(2)}(Q_{km})^{(3)}.\\
L_{GW}&=&\dot M_G=-\frac{1}{5}(Q_{ij})^{(3)}(Q_{ij})^{(3)}\label{eqn:eloss}.
\end{eqnarray}
To justify the scheme, we compare ansatz (\ref{radreaction}) with (\ref{eqn:jloss}) in Fig. \ref{fig:internal} to ensure internal consistency and (\ref{eqn:eloss}) with results from \cite{duezlgw} in Fig. \ref{fig:elosscomp}. The evaluation of the fifth quadrupole derivative is described in section \ref{sec:numerics}.\\
The gravitational wave signal is extracted in the TT-Gauge
\begin{equation}
h^{TT}_{ij}(t)=\frac{2}{r}P_{ijkl}\ddot Q_{kl}(t-r)
\label{GWextraction}
\end{equation}
where $P_{ijkl}=(\delta_{ik}-n_in_k)(\delta_{jl}-n_jn_l)-1/2(\delta_{ij}-n_in_j)(\delta_{kl}-n_kn_l)$ is the transverse-traceless projection tensor and $n_i=x_i/|\mathbf{x}|$.\\
Due to parity, the mass-current quadrupole and the octupole moment vanish in the systems considered in this paper. Otherwise we would also have to consider these terms.

\section{Relaxation scheme}
\label{app:relax}

In order to start the simulation with well-defined initial conditions, we need a scheme that produces a binary configuration with a preselected velocity field. In our case, we considered synchronized binaries, i.e. binaries with a velocity field
\begin{equation}
\vec{v}=\vec{\Omega}\times\vec{r}
\end{equation}
where $|\vec{\Omega}|$ is chosen such that the attracting gravity force is balanced by the centrifugal force.\\
To achieve this goal we add a relaxation acceleration that drives the fluid into equilibrium
\begin{equation}
\vec{f}=-\frac{1}{\tau_{relax}}(\vec{v}-\vec{\Omega}\times\vec{r})
\label{velocupdat}
\end{equation}
where $\Omega$ is given by the balancing condition
\begin{equation}
\Omega=\sqrt{\frac{\int_{\text{star 1}}\dot{v_r}d^3x+\int_{\text{star 2}}\dot{v_r}d^3x}{r_{12}}}.
\end{equation}
Here $r_{12}$ is the binary separation and $\dot{v_r}=\dot{\vec{v}}\cdot\vec{e}_r$. We approximate $\dot{\vec{v}}$ by straight forward finite differencing in time 
\begin{equation}
\dot{\vec{v}}=\frac{\vec{v}(t)-\vec{v}(t-dt)}{dt}.
\end{equation}
Eqn. (\ref{velocupdat}) is equivalent of adding a braking force $\vec{f}_{br}=-(\vec{v}-\vec{\Omega}\times\vec{r})/\tau_{relax}$. Since not $v^i$ but rather $\tilde{u}_i$ are the fundamental variables, we have to update them by means of 
\begin{equation}
\tilde{u}_{i}=\tilde{u}_i(\vec{v})=(v^i+\beta^i)wu^0\psi^4
\end{equation}
to take to changes the desired effect.\\
In order to keep the binary separation at a desired value $r_{12}$, we adjust the stars' position from time to time and set the velocity field $\vec{v}$ to a synchronized one. Then, we also need to readjust the whole field and specific momentum distribution while keeping the physical velocity $v^i$ field constant. This is done by iteratively solving the field equations (\ref{psieqn})-(\ref{chieqn}) and
\begin{equation}
\tilde{u}_{i,new}=(1-\delta)\tilde{u}_{i,old}+\delta(v^i+\beta^i)wu^0\psi^4
\end{equation}
where $\delta$ is typically chosen to be $0.3$ to $0.5$ to avoid overshooting.


\begin{thebibliography}{1}
\bibitem{ligo}A. Abromovici et al., Science {\bf 256}, 325 (1992); A. Abramovici et al., Phys. Lett. A {\bf 218}, 157 (1996).
\bibitem{virgo}C. Bradaschia et al., Nucl. Instrum. Methods {\bf A289}, 518 (1990).
\bibitem{geo}J. Hough, in Proceedings of the sixth Marcel Grossmann Meeting, edited by H. Sato and T. Nakamura (World Scientific, Singapore, 1992), p. 192.
\bibitem{tama}K. Kuroda et al., in Proceedings of the International Conference on Gravitational Waves: Sources and Detectors, edited by I. Ciufolini and F. Fidecard (World Scientific, 1997), p. 100.
\bibitem{lattimer74}J. M. Lattimer, D. N. Schramm, ApJ, {\bf 192}, L145 (1974).
\bibitem{lattimer76}J. M. Lattimer, D. N. Schramm, ApJ, {\bf 210}, 549 (1976).
\bibitem{symbalisty}E.M.D. Symbalisty, D.N. Schramm, Astrophys. Lett. {\bf 22}, 143, (1992).
\bibitem{eichler}D. Eichler, M. Livio, T. Piran, D.N. Schramm, Nature, {\bf 340}, 126 (1989).
\bibitem{Meyer}B. S. Meyer, ApJ, {\bf 343}, 254 (1989).
\bibitem{rosswog99}S. Rosswog et al., Astron. Astrophys, {\bf 341}, 499 (1999).
\bibitem{freiburghaus}C. Freiburghaus, S. Rosswog, F. K. Thielemann, ApJ, {\bf 525}, L121 (1999). 
\bibitem{Duncan}R. Duncan, C. Thompson, ApJ, {\bf 392}, L9 (1992).
\bibitem{kluzniak}W. Kluzniak, M. Ruderman, ApJ, {\bf 505}, L113 (1998)
\bibitem{oohara97}K. Oohara, T. Nakamura, Prog. Theor. Phys. Suppl., {\bf 136}, 270 (1999).
\bibitem{davies}M. Davies, W. Benz, T. Piran, F.-K. Thielemann, ApJ, {\bf 431}, 742 (1994).
\bibitem{zhuge}X. Zhuge, J.M. Centrella, S. L.W. McMillan, Phys. Rev. D, {\bf 50}, 6247 (1994).
\bibitem{zhuge2}X. Zhuge, J. M. Centrella, S. L. W. McMillan, Phys. Rev. D, {\bf 54}, 7261 (1996).
\bibitem{Rasio}F. Rasio, S. Shapiro, ApJ, {\bf 438}, 887 (1995).
\bibitem{shai}S. Ayal, T. Piran, R. Oechslin, M. B. Davies, S. Rosswog, ApJ, {\bf 550}, 846 (2001).
\bibitem{faber}J. A. Faber, F. A. Rasio, Phys. Rev. D, {\bf 62}, 064012 (2000).
\bibitem{faber2}J. A. Faber, F. A. Rasio, J. B. Manor, Phys.Rev. D, {\bf 63}, 044012 (2001).
\bibitem{wilson}J. R. Wilson, G. J. Mathews, P. Marronetti, Phys. Rev. D, {\bf 54}, 1317 (1996).
\bibitem{shibataGR}M. Shibata, Phys. Rev. D {\bf 60}, 104052 (1999).
\bibitem{shibataGRhires}M. Shibata, K. Ury$\bar{\text{u}}$, astro-ph/0104409.
\bibitem{shibatauryu}M. Shibata, K. Ury$\bar{\text{u}}$, Phys. Rev. D, {\bf 61}, 064001 (2000). 
\bibitem{ruffert96}M. Ruffert, H. T. Janka, G. Sch\"afer, Astron. Astrophys., {\bf 311}, 532 (1996).
\bibitem{max}M. Ruffert, H. T. Janka, astro-ph/0106229.
\bibitem{rosswog00}S. Rosswog, C. Freiburghaus, F.-K. Thielemann, Nucl. Phys. A, {\bf 688(1-2)}, 344C (2001).
\bibitem{rossi2001}S. Rosswog, M. B. Davies, astro-ph/0110180.
\bibitem{rasiorev}F. A. Rasio, S. L. Shapiro, Class. Quant. Grav., {\bf 16}, R1-R29 (1999).
\bibitem{blanchetpointmass}L. Blanchet, B. R. Iyer, C. M. Will, A. G. Wiseman, Class. Quant. Grav., {\bf 13}, 575 (1996)
\bibitem{blanchetpointmass2}L. Blanchet, Prog. Theor. Phys. Suppl., {\bf 136},146 (1999).
\bibitem{duezQE}M. D. Duez, T. W. Baumgarte, S. L. Shapiro, Phys. Rev. D, {\bf 63}, 084030 (2001).
\bibitem{damour98}T. Damour, B. R. Iyer, B. S. Sathyaprakash, Phys. Rev. D, {\bf 57}, 885 (1998).
\bibitem{cook}G. B. Cook, S. L. Shapiro, S. A. Teukolsky, Phys. Rev. D, {\bf 53}, 5533 (1996).
\bibitem{cookGR}G. B. Cook, S. L. Shapiro, S. A. Teukolsky, ApJ, {\bf 398}, 203 (1992).
\bibitem{shibatahij}M. Shibata, K. Ury$\bar{\text{u}}$, gr-qc/0109026.
\bibitem{flanagan}E. E. Flanagan, Phys. Rev. Lett., {\bf 82}, 1354 (1999).
\bibitem{Benz}W. Benz, in The numerical modelling of Nonlinear Stellar pulsations: Problems ans Prospects, ed. J. R. Buchler (Dordrecht: Kluwer), 269 
\bibitem{Monaghan}J. J. Monaghan, Ann. Rev. Astron. Astrophys., {\bf 30}, 543 (1992) 
\bibitem{lombardi}J. C Lombardi, A. Sills, F. A. Rasio, S. L. Shapiro, J. Comp. Phys., {\bf 152}, 687 (1999).
\bibitem{MoGi}J. Monaghan, R. Gingold, J. Comp. Phys., {\bf 52}, 374 (1983).
\bibitem{MoMo}J. Morris, J. Monaghan, J. Comp. Phys., {\bf 136}, 41 (1997).
\bibitem{siegler}S. Siegler, H. Riffert, ApJ, {\bf 531}, 1053 (2000).
\bibitem{numrep}W. H. Press, S. A. Teukolsky, W. T. Vetterling, B. P. Flannery, Numerical Recipes in Fortran 77/90, (Cambridge University Press: Cambridge) (1999).
\bibitem{hockney}R. W. Hockney, J. W. Eastwood, Computer Simulation Using Particles (Institute of Physics: London), (1994).
\bibitem{Hess}R. Hess, Dynamically Adaptive Multigrid on Parallel Computers for a Semi-Implicit Discretization of the Shallow water equations, (GMD Forschungszentrum Informationstechnik GmbH: St. Augustin) (1999).
\bibitem{BDS}L. Blanchet, T. Damour, G. Sch\"afer, MNRAS, {\bf 242}, 289 (1992).
\bibitem{baumgarte}T. W. Baumgarte, G. B. Cook, M. A. Scheel, S. L. Shapiro, S. A. Teukolsky, Phys. Rev. D, {\bf 57}, 7299 (1998).
\bibitem{wilsonshort}P. Marronetti, G. J. Mathews, J.R. Wilson, Phys. Rev. D, {\bf 58}, 107503 (1998).
\bibitem{baumgartediffrot}T. W. Baumgarte, S. Shapiro, M. Shibata, ApJ, {\bf 528}, L29 (2000).
\bibitem{duezlgw}M. D. Duez, T. W. Baumgarte, S. L. Shapiro, M. Shibata, K. Ury$\bar{\text{u}}$, gr-qc/0110006.
\bibitem{kalogera}V. Kalogera, D. Lorimer, astro-ph/9907426.
\bibitem{MTW}C. W. Misner, K. S. Thorne, J. A. Wheeler, Gravitation, (San Francisco: W. H. Freeman and Co.) (1973).
\bibitem{RS2}F. A. Rasio, S. L. Shapiro, ApJ, {\bf 432}, 242 (1994).
\bibitem{faberconf}J. Faber, F. A. Rasio, in Astrophysical Sources of Gravitational Radiation, ed. J. M. Centrella (AIP Press).
\bibitem{shen}H. Shen, H. Toki, K. Oyamatsu, K. Sumiyoshi, Prog. Theor. Phys. {\bf 100}, 1013 (1998).
\bibitem{shibatacfc}M. Shibata, T. W. Baumgarte, S. L. Shapiro, Phys. Rev. D, {\bf 58}, 023002 (1998).
\bibitem{thorne80}K. S. Thorne, Rev. Mod. Phys, {\bf 52}, 299 (1980).


\end{thebibliography}
\end{document}